\documentclass{article}

\usepackage{arxiv}
\usepackage{cite}
\usepackage{amsmath,amssymb,amsfonts}
\usepackage{algorithmic}
\usepackage{graphicx}
\usepackage{textcomp}
\usepackage{xcolor}
\def\BibTeX{{\rm B\kern-.05em{\sc i\kern-.025em b}\kern-.08em
    T\kern-.1667em\lower.7ex\hbox{E}\kern-.125emX}}

\usepackage{color}
\definecolor{Orange}{rgb}{1,0.5,0}
\definecolor{Red}{rgb}{0.8,0,0}
\definecolor{LightBlue}{rgb}{0.2,0.5,1}

\usepackage{bm}

\begin{document}

\title{Acoustic Environment Transfer for \\Distributed Systems
}
\author{
  Chunheng Jiang \\
  Department of Computer Science\\
  Rensselaer Polytechnic Institute, Troy, NY, USA\\
  \texttt{ jiangchunheng@gmail.com} \\
  \And
  Jae-wook Ahn \\
  IBM Research\\
  Yorktown Heights, NY, USA \\
  \texttt{jaewook.ahn@us.ibm.com} \\
  \AND
  Nirmit Desai \\
  IBM Research\\
  Yorktown Heights, NY, USA \\
  \texttt{nirmit.desai@us.ibm.com} \\
  
}


\maketitle

\thispagestyle{plain}
\pagestyle{plain}

\begin{abstract}
Collecting sufficient amount of data that can represent various acoustic environmental attributes is a critical problem for distributed acoustic machine learning.  Several audio data augmentation techniques have been introduced to address this problem but they tend to remain in simple manipulation of existing data and are insufficient to cover the variability of the environments.  We propose a method to extend a technique that has been used for transferring acoustic style textures between audio data.  The method transfers audio signatures between environments for distributed acoustic data augmentation.  This paper devises metrics to evaluate the generated acoustic data, based on classification accuracy and content preservation.  A series of experiments were conducted using UrbanSound8K dataset and the results show that the proposed method generates better audio data with transferred environmental features while preserving content features.
\end{abstract}


\section{Introduction}

In distributed computing, insuring proper training data for machine learning is an important issue.  Even for the general machine learning, it is a well-known challenge  to acquire sufficient amount of data and avoid overfitting by maximizing the generalizability of the data.  The distributed environments makes this problem even more difficult, as the individual environment may have different data distribution to be included in the models, while they share common attributes at the same time.  

Distributed acoustic machine learning models are not exceptions.
Among the various problem domains in distributed computing, acoustic machine learning  is useful in various applications, including classification of different kinds of sounds in city streets \cite{salamon2014dataset}, detecting anomalies in industrial settings \cite{ahn2019acoustic,koizumi2020description}, identifying health problems by observing body sounds \cite{syed2007framework}, and tracing species in danger by tracking their sounds \cite{mcdonald2006biogeographic}.  

In addition to the variability of the problem domain, the training data for acoustic machine learning models are sensitive to environmental changes.  The same object can create different acoustic signatures in different environments.  For example, a motor can produce different sounds depending on the location where it is installed in a factory and when the sounds are captured during a day. Even in the same location, the direction or distance from sound capturing devices (i.e. microphones) can result in different sounds.  If one wants to detect anomalies of motors by observing their sounds, they need to consider all these possible variations resulted from different environments.  The most straightforward way may be to manually collect as much data as possible that can represent all the possible situations.  However, it is an unrealistic assumption in many problem domains where it is not allowed to capture audio samples in many locations for a long period of time.  In fact, when one needs to collect abnormal sounds, it is difficult to expect when the anomalies would happen and the frequency is usually very low.

In order to address this problem, we need an efficient method to generate acoustic data adapted to the environmental changes that can complement the lack of data in multiple environments.  Traditionally, the machine learning community has devised various methods to face the overfitting issue, such as regularization \cite{kukaka2017regularization}, transfer learning\cite{pratt1993discriminability,pan2009survey}, and data augmentation \cite{shorten2019survey}.  Transfer learning has been acknowledged to be an effective method to expand the amount of training data by reusing a pre-trained model 
and transferring knowledge learned from one environment to another as a starting point \cite{olivas2009handbook,goodfellow2016deep}. Data augmentation is a suite of techniques that enhance the size and quality of training datasets such that better deep learning models can be built using them \cite{shorten2019survey}.

This paper attempts to use a style transfer based data augmentation technique to transfer environmental features to generate new acoustic data. Several data augmentation techniques for audio data have been introduced.  They range from simple noisy injection to more sophisticated room simulation approaches.  These simple data augmentation techniques may be suitable to specific use cases where simple variations of training data would be sufficient but they are not able to meet our requirement: newly generated data should adapt to a new environment while keeping the nature of the acoustic features of objects of interests.

Style transfer was originally developed for generating a new image \cite{gatys2015neural} that resembles the texture of a ``style'' image while maintaining the structure of objects in a ``content'' image.  A well-known example is to convert a photo (content) with the brush touch of Vincent van Gogh (style).  The resulting image looks like a painting of the content object drawn with the brush and the technique used in the style image.  This style transfer techniques have tried other problem domains such as text and audio.  When applied to text data, original texts are transformed between different times \cite{jhamtani2017shakespearizing}, sentences are revised \cite{muller2017sequence}, or the sentiments appearing in texts are transferred \cite{li2018delete}.  Recently, this technique has been used for transferring styles of a source audio \cite{grinstein2018audio} and generating a new audio that resembles styles or textures of the original style audio data.  However, the attempts mostly aimed to generate new audio samples that copy the acoustic texture of style audios and were limited to limited utilities such as varying the timbre of musical instruments. We believe that acoustic style transfer can be effectively used for data augmentation in distributed environments, by transferring  acoustic environment style to another environment, and address the data sparsity problem.  
This paper provides an experiment that attempts to 
find the optimal way for style transfer based acoustic data augmentation.
 This paper proposes a pair of evaluation criterion 
on the quality of style transfer from the needs of data augmentation. 
One is prediction accuracy based, another one is similarity or distance based.
We implement a wide random convolutional neural network as the neural style transfer model,
and carry out a series of experiments on UrbanSound8K. The style transfer
model produces better style transfer than the baseline sound mixing approach.
Meanwhile, the content are well-preserved by our transfer model. 
Also, some related hyper-parameters are examined, and both the loss function and
the architecture of the transfer model affect our transfer performance.






\section{Related Work}



With respect to the objective of this study, we found several audio data augmentation techniques from the ones that focus on a single feature to the ones that consider more complicated aspects of the sound generation. We review each of them in this section and introduce the style transfer techniques that are more directly related to our method.

\subsection{Simple acoustic data augmentation}

Simpler acoustic data augmentation techniques that manipulate a specific dimension have been widely used for neural network based speech recognition \cite{schluter2015exploring,mcfee2015software,parascandolo2016recurrent,park2019spec}.  Following is a list of the techniques and a brief their descriptions.

\begin{enumerate}
    \item Noise Injection -- Add random noises into original sounds.
    \item Shifting Time -- Shift audio to left or right with a random seconds.  Silence is added to the shifted space.
    \item Changing Pitch -- Change pitch randomly.
    \item Changing Speed -- Stretch time series by a fixed rate.
\end{enumerate}

Salamon et al. \cite{salamon2017deep} has reviewed these techniques and discussed their shortcomings despite their wide availability.  Moreover, simple audio manipulation is not able to separate content and background audio features and cannot variate the audio data that could sound differently under various settings.  

Another approach that is fits better to our motivation is \textit{room simulation}. It can simulate recordings of arbitrary microphone arrays within an echoic room. It supports research related to developing and experimenting with multichannel microphone arrays and higher order ambisonic playback. it models both specular and diffuse reflections in a shoebox type environment \cite{ustyuzhaninov2016texture,schimmel2009fast}. Even though room simulation is an interesting approach and it takes multiple microphones into consideration, it still assumes a single environment (a room) and considers limited number of ad-hoc features: e.g., distance between sound source and mics only.

\subsection{Image and acoustic style transfer}

Image stye transfer \cite{gatys2015neural,gatys2015texture,gatys2016image} generates new images by transferring style features from one image (called style image) to another (called content image).  The visual styles such as brush strokes and surface texture are extracted from the style image and applied to the content image using a neural network.  The texture extraction is done by using convolution filters and gram matrix and the transfer is completed using a neural network that minimizes an error function estimating the difference between the original and the transferred content.  This technique has been able to convert photos that contain real-world objects (e.g., a building) to a picture that resembles that the objects are painted with the materials or the techniques (e.g., Vincent van Gogh painting style) used in the style image, and were able to create images that look quite realistic in the sense of the visual styles.

Acoustic style transfer is motivated from the original image style transfer.  Based on the fact that acoustic signals can be easily converted to visual images such as spectrograms and the spectrograms can represent the audial features visually, they applied the image style transfer techniques onto the acoustic spectrogram \cite{grinstein2018audio}.  In the spectrogram, the audio signals are converted/visualized as heat-maps where the horizontal axis represent frames (or time), vertical axis represent frequencies, and the color or intensity of the pixels represent the strength of the signal in the given time and frequency.  The existing acoustic style transfer approaches use a similar method to extract the style features but adapted to audio by defining a vertically narrow convolutional feature maps or adopts an auto-encoder based approach \cite{ramani2018autoencoder}.  With the maps, the transfer model can capture acoustic signature across multiple frequency, ranging within a short amount of time.  The style features are obtained by feeding style sound to pre-defined VGG models\cite{simonyan2014very} or a random noise layers \cite{ustyuzhaninov2016texture}, and then calculating a gain matrix of the activations.  The use of random noise as filters was proven to work with image style transfer, as the textures share visual elements with simple random noises.

\section{Acoustic Environment Transfer using Style Transfer Method} \label{sec:system}

\begin{figure*}[t]
\centering
\includegraphics[width=0.9\textwidth]{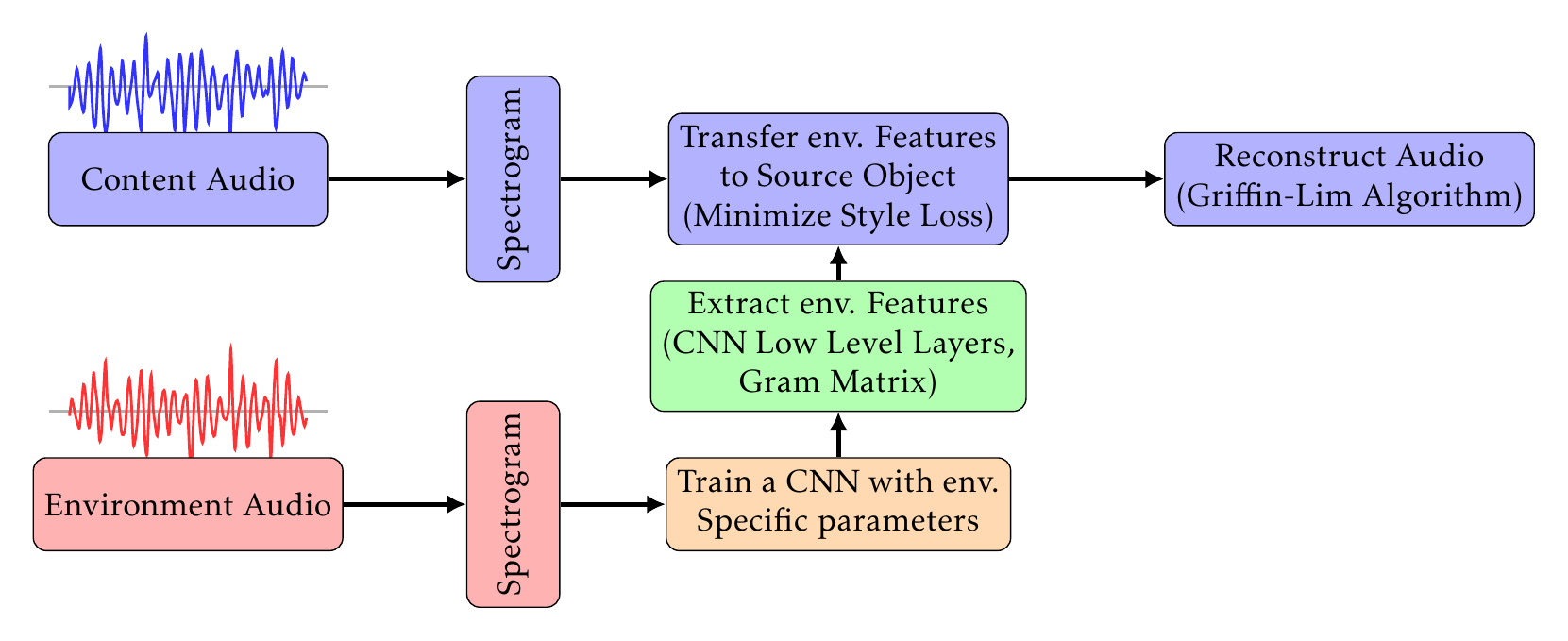}
\caption{\textsf{Environment transfer for audio framework: the environment audio and content audio are transferred to spectrogram and each feature is extracted using a CNN.  The environment feature is transferred to the content audio so that the resulting audio can represent the change of the content audio in a different environment.  The new strategy (the new convolutional filter configuration) to enhance the method for environment audio takes places in the lower row of the diagram.}}
\label{fig:diagram}
\end{figure*}

We implemented an acoustic environment transfer system based on the common style transfer method \cite{grinstein2018audio,gatys2016image}.  The main difference between the previous work (style transfer) and the current work (environment transfer) is what is transferred from the source to the target sound.  The style transfer focuses on the features such as visual textures or audio timbre that changes the feeling of the generated the data.  On the other hand, the environment transfer focuses on transferring acoustic environment where a content object generates sound in different environments. If the environment is different the sound that the object makes will be different.  This acoustic environment will not only include simple background sound but also include specific environment features that could affect the content sound such as echo or reverb.  We hypothesize that these environmental features could be captured as \textit{acoustic styles}.  Therefore, while we adopt the core idea from the acoustic style transfer, we need to adapt the feature extraction procedure that can represent different environments.  

\begin{figure}[t]
    \centering
    \includegraphics[width=0.5\textwidth]{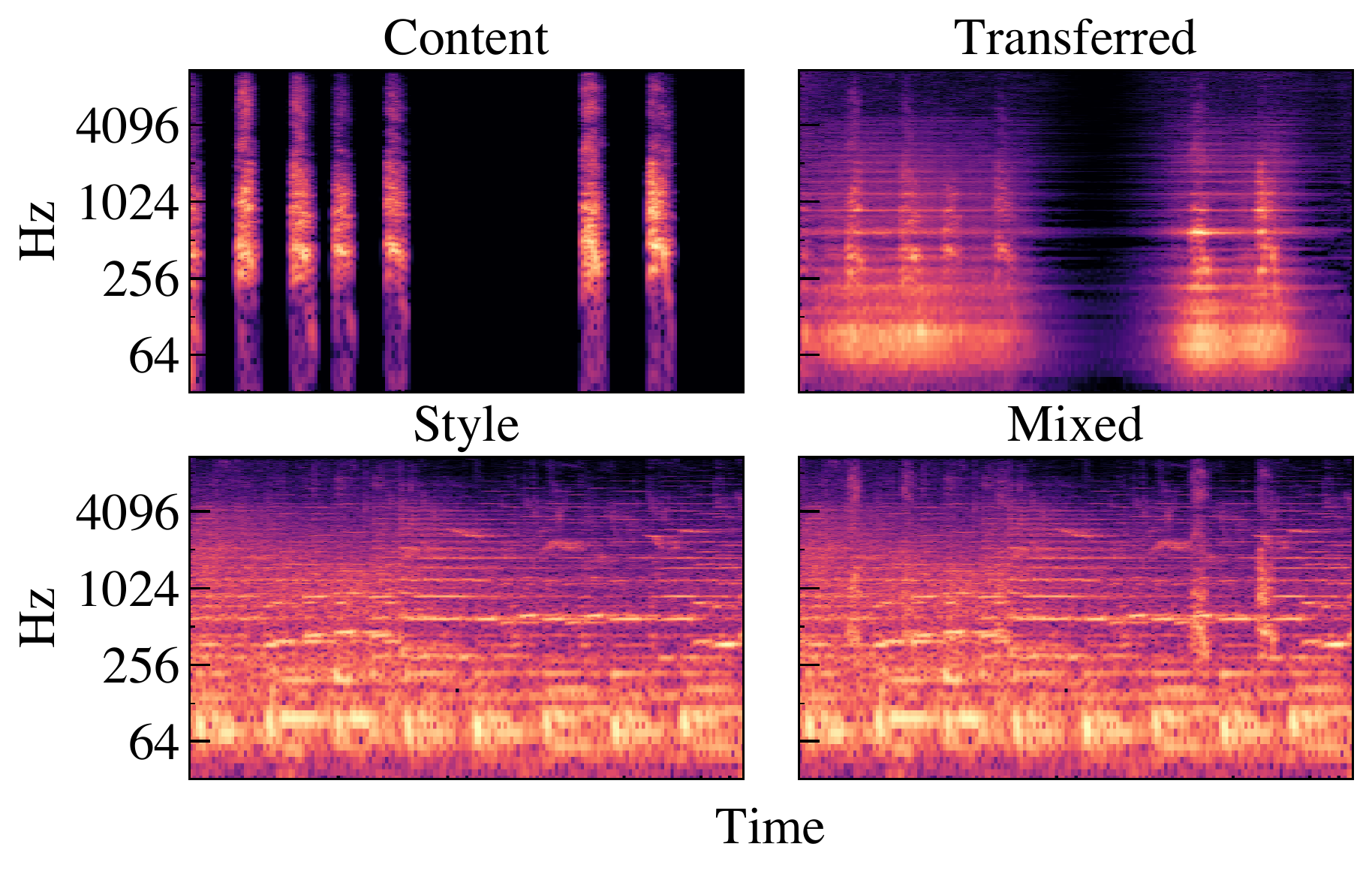}
    \caption{Comparison between style transfer and audio mixing: with a content audio (dog barking) and a style audio (street music) (left column), example spectrograms of our neural style transfer model generation (upper right) and the simple sound mixing (lower right) are shown.  The transferred sound shows the six acoustic signals in a column format (six dog barks) and the attributes from the style audio (repeatedly playing street music in the background).  The mixed audio shows the content signal is simply added on top of the background. }
    \label{fig:example}
    \end{figure}

Figure~\ref{fig:diagram} shows the overall architecture of acoustic environment transfer.  In order to extract two features: (1) environment features and (2) content features, the source audio files are translated to spectrograms.  
Spectrogram is a 2D visual representation of frequencies of 
a given signal with time, while the color represents the magnitude or amplitude (see Figure\ref{fig:example}).
Spectrograms contain rich information of the audio signal, 
including the frequency, the time and the amplitude. 
Therefore, it is often used as a visualization tool for representing audio signals such as music and speech.
Also, it is widely used 
for audio classification\cite{gemmeke2017audio,hershey2017cnn} and style transfer\cite{grinstein2018audio,ramani2018autoencoder}.


The spectrograms are converted to low-level features by the convolution layer.  There have been discussions how to define the convolution layer: use pre-defined model such as VGG-16 or random noise.  VGG-16 has been successfully used by image style transfer \cite{simonyan2014very,gatys2016image} but \cite{grinstein2018audio} argued that it was not as robust when applied to acoustic style transfer, because it was trained with images while spectrograms show different visual elements such as abstract-looking waves and shades and even reported random noises produce similar results.  This claim also holds for transferring specific images where the styles are mostly material textures that are visually similar to random noises \cite{ustyuzhaninov2016texture}. 


In order to implement acoustic style transfer more adapted to environment transfer, we propose variable convolutional filter configurations.  The previous work defined the filters that can capture the features within a short frame.  From our experience with the use cases such as (Section~\ref{sec:use-case}), we learned that it is critical to find out the optimal window size (frame size) for sampling sounds and training models, and we believe it is still true for capturing environmental sound features.  If the windows size is too large, then the data can include redundant or repeating features that would lead to performance degradation and if the size is too small it will lose critical features including temporal patterns that spans over the predefined window.  Therefore, in this study we vary the configuration of the convolution filters with respect to the frame size and attempt to find out the best set-up for environment sound transfer.

After the low-level features are collected through the convolutional network, they are converted to gram matrices and are used for the transfer stage. Given the representation $\bm x$ of an input audio signal (waveform or spectrogram), a convolutional neural network architecture is used to extract statistics that characterize stationary sound textures.
Let $F_{\ell}=[f_{{\ell},k}]_{k=1}^{N_\ell}$ be activation vector 
on layer $\ell$ with $N_\ell$ nodes. 
Following the practice in \cite{gatys2016image}, 
we used the Gram matrix $G_{\ell}=F_\ell^T F_\ell$ as the style loss statistics,
and minimized a two-fold loss function
\begin{equation}
\mathcal L(\bm x,\bm x_c,\bm x_s)=\alpha\mathcal L_c(\bm x,\bm x_c)+\mathcal L_s(\bm x,\bm x_s)
\end{equation}
to transfer the target style, 
where $\bm x_c$, $\bm x_s$ and $\bm x$ are the content, the style and the generated signals, respectively;
$\mathcal L_c(\bm x,\bm x_c)=\sum_{\ell\in \mathcal C} \|F_\ell(\bm x)-F_\ell(\bm x_c)\|_2^2/N_\ell$
and $\mathcal L_s(\bm x,\bm x_s)=\sum_{\ell\in \mathcal S} \|G_\ell(\bm x)-G_\ell(\bm x_s)\|_F^2/N_\ell$
are the content and style loss, respectively; 
$\alpha>0$ controls how much penalty will be exercised over the deviation of the generated audio from the content loss,
$\mathcal C$ and $\mathcal S$ are the indices for content and style layers.
The size of $\mathcal S$ varies between different neural network architectures. 
For example, $|\mathcal S|=5$ in VGG-16, $|\mathcal S|=8$ in SoundNet\cite{aytar2016soundnet} and 
$|\mathcal S|=1$ in the wide-shallow-random network by Ulyanov and Lebedev\cite{ulyanov2016audio}.
In our simulation, we have $|\mathcal S|=1$.

The transfer model generates a new spectrogram image, from which we reconstruct the missing phase
and generate a waveform audio with the Griffin-Lim algorithm\cite{griffin1984signal}.

Figure~\ref{fig:example} shows example spectrograms of a style audio, content audio, style-transferred audio (from the style to the content), and sound mixing audio of the style and content audio \ref{sec:condition1}.  The spectrogram generated by our proposed method is clearly distinguished from the others.  It captures the features from both the original style and the content audio (left column) while not being dominated by either of them.  Meanwhile, the audio generated by audio synthesis is just an addition of two separate audio.  We will discuss more about the difference in the later sections.

Our system is built upon the training of three convolutional neural networks:
a {\it style transfer model} to transfer the source style into the target,
an {\it audio classifier} to predict the class of the sound clips, 
and an {\it AutoEncoder} for embedding features.
Both the classifier and the AutoEncoder are trained to evaluate the quality of our style transfer model.

\subsection{Style transfer network}
The style transfer model is a shallow, wide, random single-layer neural network with 4096 convolutional filters.

\subsection{Audio classifier} 
The model contains four convolutional layers of $3\times 3$ 
(with ReLU activation functions, 32, 32, 64, and 64 channels, respectively), two pooling layers, 
followed by a flatten layer of size 6656 and two fully connected layers. 
Three dropout layers are applied after the pooling layers and the first dense layer 
with a drop rate of 0.15, 0.2 and 0.5, respectively.

\subsection{AutoEncoder}
The AutoEncoder contains an encoder and a decoder, 
the encoder maps the input to a latent space, 
and the decoder maps the latent representation to the input space.
The encoder has two convolutional layers of $3\times 3$ (with ReLU activation functions, 16 and 8 channels, respectively),
and each of which is followed by a pooling layer. The decoder  contains three convolutional layers (with two ReLU activation functions and a softmax, 8, 16 and 2 channels, respectively). The input shape is of $60\times 32\times 2$, 
and the latent dimension is $15\times 8\times 8$.

\section{Example Use Case} \label{sec:use-case}

Among various cases that need to train a model with a limited amount of training dat, we present an example use case scenario of the acoustic environment transfer in order to help the understanding of readers.  In an industrial manufacturing site, an accident can happen any time and it is challenging to manually monitor the situation 24 hours a day, depending on the nature of the manufacturing.  Some sites would require limited number of staffs during and after work hours.  Moreover, visual inspection can be inherently impossible if the accident happens inside of facilities.  Automated acoustic monitoring and inspection can be effective in those sites and acoustic anomaly detectors or audio classifiers are need to be trained.  However, it is not trivial to collect the training data, especially if it is anomaly data because anomaly itself assumes that it does not happen frequently.  It will increase the performance of the anomaly detector if we can generate new data based on the limited training data by allowing the variability of the anomaly state within the problem space.  Conventional data augmentation technique will be helpful by simply manipulating the pitch or length of the sounds but we still need to incorporate another variable: acoustic environment.  The importance of the environmental sound is more important in the aforementioned industrial sites, where background noise exists almost all the time that can continually change over time and can introduce abrupt acoustic change depending on manufacturing process changes.  These background sound variation can greatly affect the performance of the acoustic models, especially when the training data is limited.  We could improve the models if we can transfer the various environment sounds (relatively easily collected) to the limited number of anomaly sample and the resulting sounds will be able to cover anomalies happening in multiple situations.  
\section{Study Design}
\subsection{Research Questions}

We ask two questions in this study.  We hypothesize that the acoustic style transfer method can be successfully used to transfer acoustic environment features between distributed environments and augment data for application areas such as sound classification and anomaly detection.  We have discussed simpler data augmentation methods but they simply manipulate the source data with respect to several basic features and there is no attribute that could be used to represent different environments.  Rather than those data augmentation methods, we adopted an audio mixing method that combines the content and environment audio.  It is a common method to combine two audios.  We will discuss about it in more detail in Section~\ref{sec:condition1}.

The second question is to discover what is a better way between environment transfer methods.  We configured our system either: (1) with same settings with the default `style transfer' system (Section~\ref{sec:condition2}) (2) with variable convolutional filter sizes in order to better capture the sound signatures from various environments (Section~\ref{sec:condition3}).  We would like to prove if this assumption leads to better results.

\begin{enumerate}
    \item RQ1 -- Is the acoustic environment transfer method better than the simple baseline audio mixing in order to adapt content sound in distributed environments?
    \item RQ2 -- Is the optimal acoustic environment transfer method optimized for environment transfer better than the existing acoustic style transfer that is generic to audio data manipulation rather than environment transfer?
    
\end{enumerate}
\subsection{Dataset}
UrbanSound8K\cite{salamon2014dataset} is used in our experiments. 
It is comprised of 8732 sound clips of up to 4 seconds in duration 
taken from the field recordings. The clips span ten environmental sound classes: air conditioner, car horn, children playing, dog bark, drilling, engine idling, gun shot, jackhammer, siren, and street music. 
 Also, each sound clip is assigned a subjective salience label in order to indicate a given sound is a foreground (salience=1) or background one (salience=2).  This is an interesting attribute particularly for environmental sound analysis.  Common audio name labels does not deliver enough information to determine if the sound is foreground or background.  For example, `Siren' sounds could be considered either background sound (e.g., an ambulance is going by in a distant location while a dog is barking) or a foreground sound (e.g., the sound clip is about an ambulance sound recorded directly). We found that the background sounds are good candidates that include environmental sounds (e.g., background air conditioner working sound or background street music) while the foreground sounds are good content sounds (e.g., drilling or gun shot sounds in the foreground) to which the environment sounds are transferred. We summarize related statistics in Table~\ref{tb:stat}.

Additionally, each sound clip is tagged with its recording file name, 
and annotated with the start and end time in the original recording sound.
Because of the variation presented in the time duration, 
we concatenate these the short clips according to the time annotations, 
and produce 933 foreground recordings, 404 background recordings.
From which, we pair 1000 sound clips of 4-second length from different classes, 
where the foreground sound clip as the target content and 
the background sound clip as the target style.
These 1000 pairs are used to evaluate our style transfer model.
Our audio classifier is trained on 80\% randomly selected sound clips,
and tested on the remaining 20\% clips. The model selection is made based on 10\% validation set in the training set to avoid overfitting. Similarly, we train the AutoEncoder network with 80\% randomly selected clips as the training set, and the remaining 20\% as the testing set.
\begin{table}[ht]
\centering
\caption{Statistics of UrbanSound8K by classes and salience.}
\label{tb:stat}
\begin{tabular}{llll}
Class & Foreground & Background & Total\\
\hline
Air Conditioner & 569 & 431 & 1000\\
Car Horn & 153 & 276 & 429\\
Children Playing & 588 & 412 & 1000\\
Dog Bark & 645 & 355 & 1000\\
Drilling & 902 & 98 & 1000\\
Engine Idling & 916 & 84 & 1000\\
Gun Shot & 304 & 70 & 374\\
Jackhammer & 731 & 269 & 1000\\
Siren & 269 & 660 & 929\\
Street Music & 625 & 375 & 1000\\
\hline
\end{tabular}
\end{table}

\subsection{Experiment conditions}

We define three conditions: sound mixing without using style transfer, default style transfer without convolutional filter variation, and style transfer with the filter variation.  

\subsubsection{Sound mixing} \label{sec:condition1}
This method overlays a sound on top of another.  It may be the first method one can intuitively consider when trying to adapt a content to a new environment.  However, simple mixing does not distinguish the environment and the content and the limitation clearly appears in the result sound.  In addition to the simple subjective observation, we will explore it quantitatively in the experiment.

\subsubsection{Default acoustic style transfer} \label{sec:condition2}
This is the default style transfer method described in Section~\ref{sec:system}.  The same set-up is used as in \cite{grinstein2018audio} that was built for generic sound style transfer, rather than specific to environmental sounds.

\subsubsection{Acoustic style transfer with variable convolution filter sizes} \label{sec:condition3}

We compare different filter sizes, hoping they can capture more features specific to environmental sounds, rather than acoustic textures that would define simple timbre variations.

\subsection{Evaluation metrics}
It is hard to define the notion of content and style in audio style transfer, 
and no universal agreement on the definition of content and style has been reached, 
even in well-developed visual style transfer.
However, there are some widely accepted principles, e.g.,
the visual style refers mostly to the {\it space-invariant intra-patch statistics}, i.e., to the
texture at several spatial scales, and to the distribution of colors a.k.a. the color palette;
the visual content represents the {\it broad structure of the scene}, that is, 
its semantic and geometric layout\cite{grinstein2018audio}. 
For acoustic scenarios, it usually strongly depends on the context.

Therefore, it is not surprising that the previous studies relied on qualitative and subjective assessments \cite{grinstein2018audio}. 
However, the goal of environmental audio transfer in this study is specifically focusing on data augmentation across environments: generate new acoustic data adapted from one environment to another. To make the evaluation easier, 
we develop a set of evaluation criterion on generic environmental acoustic style transfer 
from the perspective of data augmentation.

The purpose of data augmentation is to generate some rare data that are not observed in the 
given dataset. Conventional data augmentation approaches are based on random perturbations over existing samples.
The neural style transfer based framework is proposed as an alternative. The framework generates 
new data from pairs of target audio clips, one contributes the content, 
another one supplies the style.

There is very few previous work on the evaluation of style transfer.
Fu et al.\cite{fu2018style} defined two evaluation metrics for {\it style transfer in text}: 
the {\it transfer strength} indicates how the newly generated data can be accurately classified as the target \cite{fu2018style,shen2017style}, and the {\it content preservation rate} measures the similarity between
the source embedding and the target embedding. Text style transfer is similar to style transfer in speech\cite{jia2018transfer}, and the preserved content should be meaningful and recognizable. 
Therefore, it requires high accuracy in classification.

Our goal is to ensure the semantic content well-preserved, whilst the target style is transferred in
the newly generated data. 
For the applications of classification, these transferred data should be able to 
cheat the state-of-the-art classifier $\mathcal F_0$ trained from the dataset without these instances. 
With these transferred data included, we retrain the model for a new classifier $\mathcal F_1$.
It would be more robust and have better generalization. 
Therefore, the value of the style transfer model
can be estimated using the prediction accuracy increase with the transferred data included. 
Let $P_0$ and $P_1$ be the accuracy of $\mathcal F_0$ and $\mathcal F_1$ on the same testing set,
and $P_1-P_0$ is the value of the style transfer model.
Equivalently, we can evaluate the performance of the same classifier $\mathcal F_0$ 
on the generated data from our transfer model, and the synthesized data from the simple sound mixing.
Given the same pairs of content and style audios, 
the transfer model generates $\mathcal D_t$ and the simple mixing model produces $\mathcal D_m$,
Suppose $\mathcal F_0$ gives the prediction accuracy of $P_t$ and $P_m$, respectively.
The lower of the prediction accuracy $P_t$ relative to $P_m$, the better of the style transfer model. 
The idea forms a contrast to the proposed transfer strength in text style transfer\cite{fu2018style}.
The text style transfer requires the generated data recognizable, 
and it is very sensitive to style noise.
But our purpose is different, we are concentrating on data augmentation. 
If we can generate ``high quality" data to beat $\mathcal F_0$,
we succeed. 

To be of high quality, we impose additional constraints over the transferred data. 
Similar to the content preservation proposed in\cite{fu2018style},
we require the content well-preserved in the transferred data.
To meet this end, we obtain the representative features from raw audio using an AutoEncoder. 
With these features, we can calculate the similarity between the generated data and source audios. 
The more similar, the more preserved and the better.







\section{Experimental Results}
We evaluate the performance of our style transfer model on 1000 pairs of foreground and background sound clips
of 4-second length from different classes in UrbanSound8K. 

To answer {\bf RQ1}, we select the simple sound mixing as the baseline,
and compare its performance with that of the style transfer model, based on our proposed evaluation criterion.

There are many hyper-parameters in the framework, some of them may greatly 
affect the performance of our style transfer model, therefore impact the comparison results.
Hence, we study the impacts from two aspects, one is the objective function, another
is the network architecture. The scalar factor $\alpha$ in the joint loss function 
and the filter size of the convolutional layer 
are selected as representative parameters to study. The tuning of these parameters can therefore answer {\bf RQ2}
for an optimal style transfer.

\begin{figure}[b]
\centering
\begin{minipage}{1\linewidth}
\includegraphics[width=\textwidth]{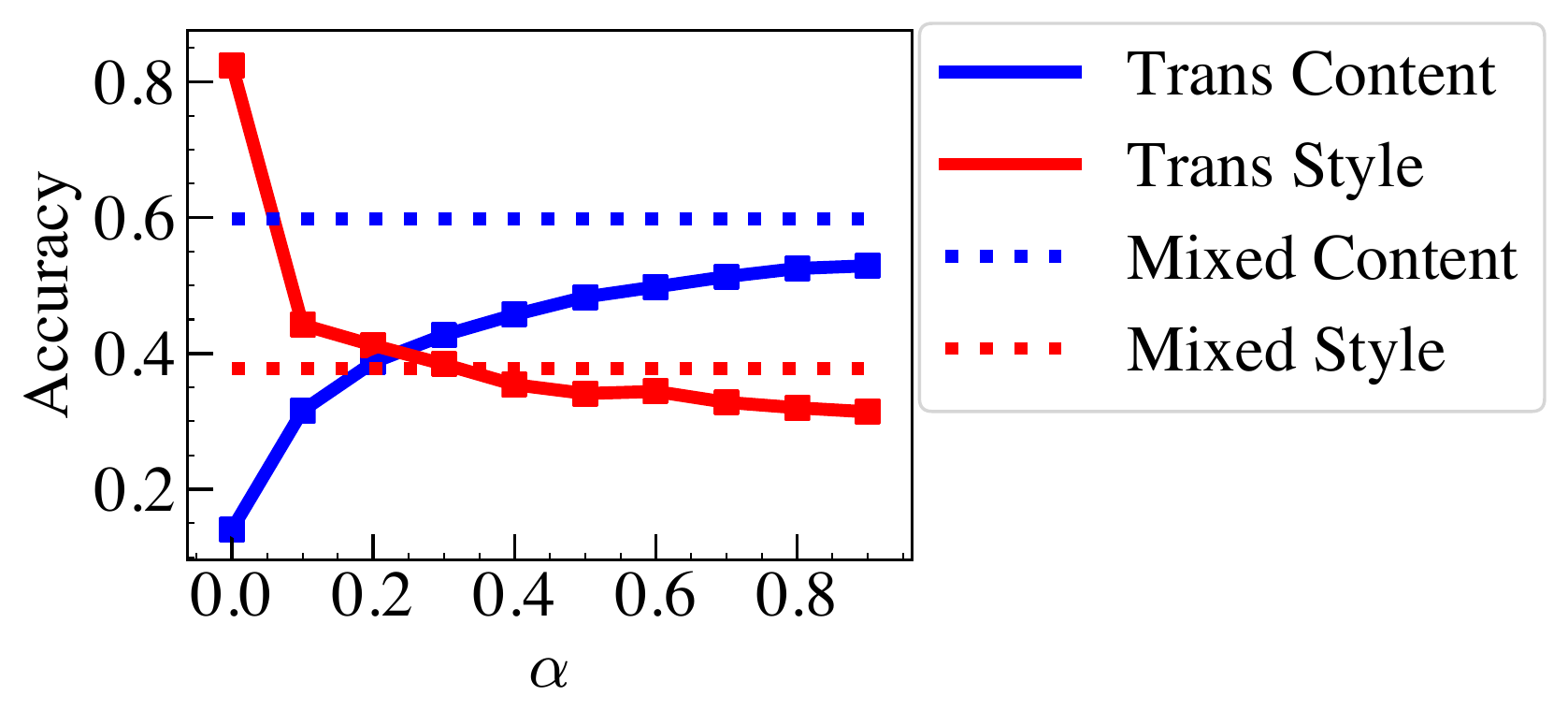}
\end{minipage}
\begin{minipage}{1\linewidth}
\includegraphics[width=\textwidth]{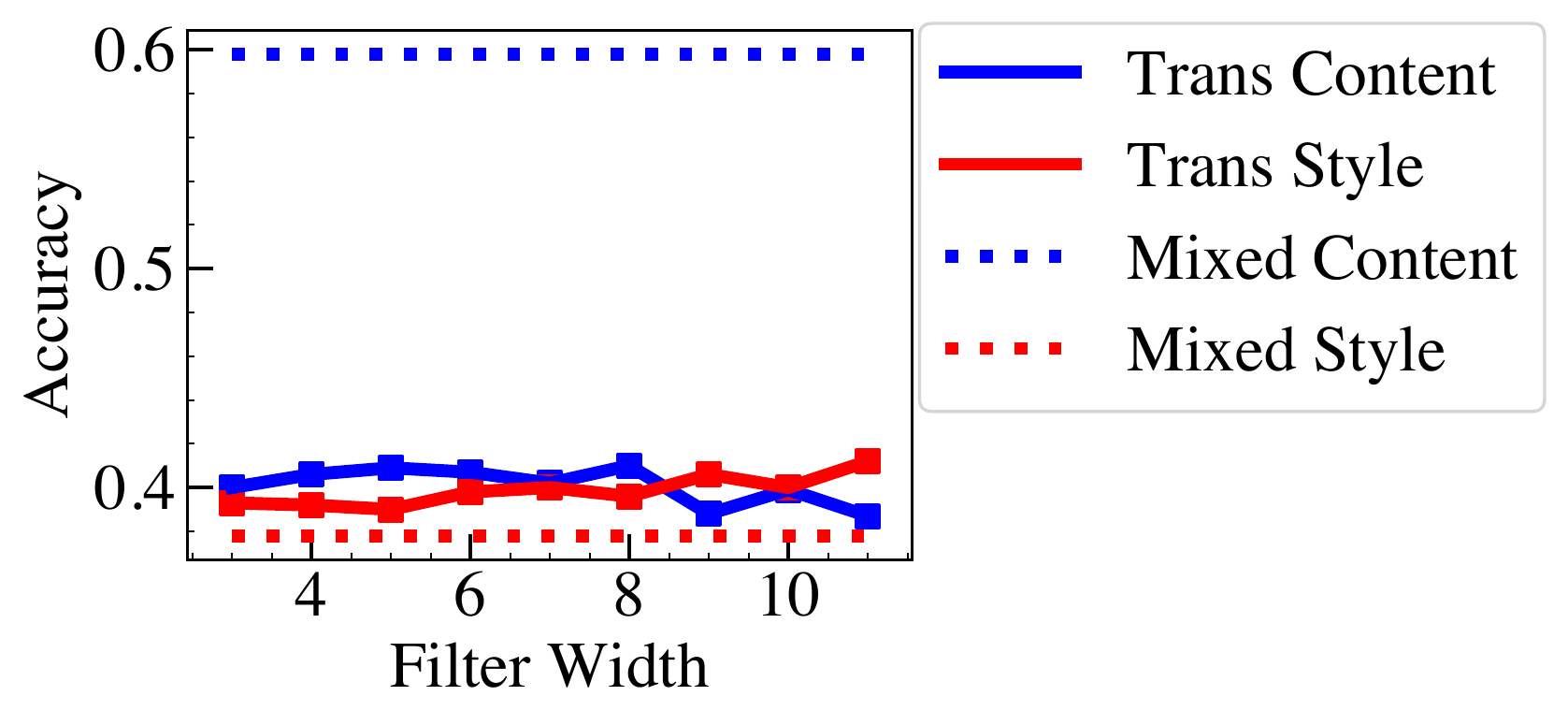}
\end{minipage}
\caption{Impacts of $\alpha$ (top) and the filter width (bottom) on the transferred data and the classifier prediction performance.}
\label{fig:impact_alpha_filter_width}
\end{figure}

According to Figure~\ref{fig:impact_alpha_filter_width}, 
as the factor $\alpha$ increases from 0 to 0.9,
the recognizing ability of the trained classifier with respect to the content 
is negatively affected by the transferred style. 
Put it in other words, the generated data losses some structure information.
Meanwhile, the target style has an obvious increasing trend in the generated data.
When $\alpha=0.2$, we achieve same the prediction accuracy on content and style. 
It indicates that the prediction uncertainty is high, we succeed to beat the classifier.
At the same time, the accuracy on the transferred data with respect to the content and style 
is less than that of the synthesized data from the baseline. The difference in accuracy 
between the baseline and the generated data from the style transfer model indicates the value of the
transfer model.

The impacts of the filter size on the prediction accuracy is very stable, either similar to 
the baseline on the content prediction, or lower than the baseline on the style prediction. 
Again, the lower the prediction, the better of the model, and the style transfer model wins out.

\begin{figure}[b!]
\centering
\begin{minipage}{.6\linewidth}
\includegraphics[width=\linewidth]{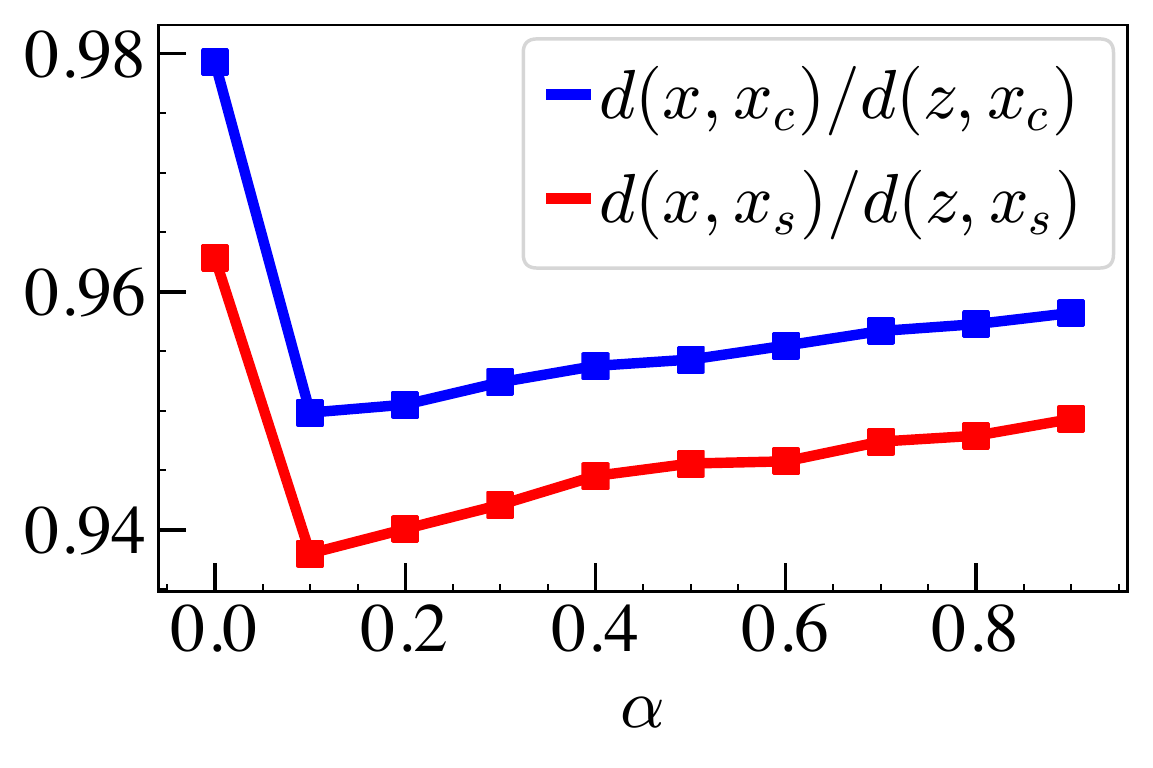}
\end{minipage}
\begin{minipage}{.6\linewidth}
\includegraphics[width=\linewidth]{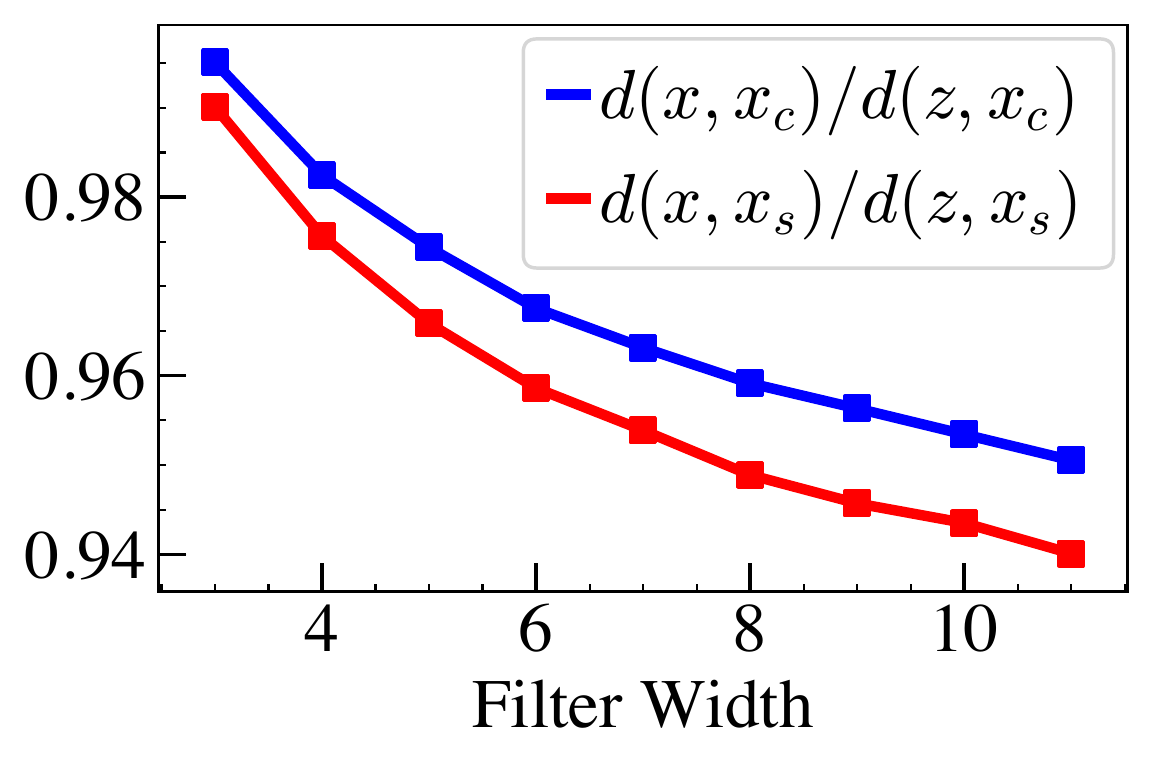}
\end{minipage}
\caption{Impact of $\alpha$ (top) and the filter width (bottom) on the transferred data 
and the distances $d({\bm x},{\bm x}_c)$ of the transferred data to the target content and the target style. 
Here, ${\bm x}_c$ and ${\bm x}_s$ are the embedding features of the content audio and
the style audio, respectively; ${\bm z}$ is the feature map of the synthesis from simple sound mixing approach, 
and ${\bm x}$ is the representative feature vector of transferred audio; $d(\cdot,\cdot)$ is the Euclidean distance.
The denominators $d({\bm z},{\bm x}_c)$ and $d({\bm z},{\bm x}_c)$ play as normalization terms. 
}
\label{fig:preservation}
\end{figure}

The preservation in content is measured with embedding features of the two involved inputs and the generated data.
With the AutoEncoder, we are able to extract an embedding feature for each input audio, then
we calculate the distance $d(\bm x, \bm x_c)$ between the generated data and the content audio, 
and the distance $d(\bm x, \bm x_s)$ of the generated data to the target style audio, 
where $\bm x$, $\bm x_c$ and $\bm x_s$ are the embedding features of the generated data, the content audio
and the style audio, respectively. The simple audio mixing is selected as the baseline approach. 
The related distances $d(\bm z,\bm x_c)$ and $d(\bm z,\bm x_s)$ are compared against those for the generated data.
As illustrated in Figure~\ref{fig:preservation}, the generated data are most similar to 
both the content and the target style audios when $\alpha=0.2$ or the filter size is large.
The distance ratios of the generated audios to the baseline are less than one, which indicates that 
the content preservation is better than the simple mixing approach.

\section{Conclusions and Future Work}
 To design a reliable evaluation metric for acoustic style transfer, 
we propose a pair of evaluation criterion to generate diverse data for the purpose of data augmentation.
One is built on the classification accuracy on the transferred data,
another one relies on the similarity between the transferred data and the input content and target style.
The classification performance on the transferred data includes the prediction accuracy on the content,
and the accuracy on the style. The generated data is expected to beat the classifier that trained from 
dataset without these transferred instances included. 
Style transfer can introduce diverse environmental scenes to the training set.
Once our model can accurately recognize these transferred data, 
it will be more robust and have a better generalization 
than the one trained from the content or style only audios.
Another objective is maximize content preservation. The input semantic content, if well-preserved
in the generated data, will keep complete in structure. 
For speech and music style transfer, the content must be strictly preserved, 
otherwise the neural transfer model 
probably generates some utter meaningless sounds\cite{mehri2016samplernn}.

Moreover, we examined two crucial hyper-parameters regarding their impacts on the quality of the style transfer
based on our proposed criterion. The penalty factor $\alpha$ over content deviation affects how much content 
is preserved in the transferred data. The higher the value, the more content preserved. 
Meanwhile, the amount of style transferred will be negatively affected. 
The trade-off is accurately depicted by the prediction accuracy change on style and content in the transferred data.
To some extend, it confirms a dependency relation between the content and style.
It indicates that the acoustic content and style are not as separable as in the visual applications\cite{gatys2016image}.

The neural style transfer model itself may exhibit another cause of the trade-off. 
Therefore, we also examined the size of the convolutional filter. 
Relatively, the filter size does not affect the transfer as much as $\alpha$.

Our style transfer experiments are built on a single layer, wide random CNN.
Therefore, the layer index set $\mathcal C$ for content equals to the index set $\mathcal S$ for style.
It causes some dependency between the style and content representations for sure.
To have a deep understanding of the relation between acoustic content and style,
we need a comprehensive study of the hyper-parameters in the neural style transfer model, 
other parameters, e.g., an architecture different from CNN, the number of layers, 
may also greatly influence the transfer quality.

\bibliographystyle{abbrv}

\end{document}